\documentclass[
reprint,
amsmath,amssymb,
aps,prl,superscriptaddress,
showkeys
]{revtex4-2}

\usepackage{graphicx}
\usepackage{dcolumn}
\usepackage{bm}
\usepackage{xcolor}
\usepackage{float}
\usepackage{soul} 
\usepackage{multirow}
\newcommand{\EE}[2]{\ensuremath{{#1}\times 10^{#2}}}

\begin{document}

\preprint{APS/123-QED}

\title{Machine-Learning-Enabled Measurements of Astrophysical (p,n) Reactions with the SECAR Recoil Separator}

\author{P. Tsintari}\email{tsint1p@cmich.edu}
 \affiliation{Department of Physics, Central Michigan University, Mt. Pleasant, MI 48859} 
\author{N. Dimitrakopoulos}
\affiliation{Department of Physics, Central Michigan University, Mt. Pleasant, MI 48859}
\author{R. Garg}
\affiliation{Facility for Rare Isotope Beams, Michigan State University, East Lansing, MI 48823}
\author{K. Hermansen}
\affiliation{Department of Physics and Astronomy, Michigan State University, East Lansing, MI 48823}
\affiliation{Facility for Rare Isotope Beams, Michigan State University, East Lansing, MI 48823}
\author{C. Marshall}
\affiliation{Department of Physics and Astronomy, Institute for Nuclear and Particle Physics, Ohio University, Athens, OH 45701}
\affiliation{Facility for Rare Isotope Beams, Michigan State University, East Lansing, MI 48823}
\author{F. Montes}
\affiliation{Facility for Rare Isotope Beams, Michigan State University, East Lansing, MI 48823}
\author{G. Perdikakis}
\affiliation{Department of Physics, Central Michigan University, Mt. Pleasant, MI 48859}
\affiliation{Facility for Rare Isotope Beams, Michigan State University, East Lansing, MI 48823}
\author{H. Schatz}
\affiliation{Department of Physics and Astronomy, Michigan State University, East Lansing, MI 48823}
\affiliation{Facility for Rare Isotope Beams, Michigan State University, East Lansing, MI 48823}
\author{K. Setoodehnia}
\affiliation{Facility for Rare Isotope Beams, Michigan State University, East Lansing, MI 48823}
\author{H. Arora}
\affiliation{Department of Physics, Central Michigan University, Mt. Pleasant, MI 48859}
\author{G.P.A. Berg}
\affiliation{Department of Physics and Astronomy, University of Notre Dame, Notre Dame, IN 46556}
\author{R. Bhandari}
\affiliation{Department of Physics, Central Michigan University, Mt. Pleasant, MI 48859}
\author{ J.C. Blackmon}
\affiliation{Department of Physics and Astronomy, Louisiana State University, Baton Rouge, LA 70803}
\author{C.R. Brune}
\affiliation{Department of Physics and Astronomy, Institute for Nuclear and Particle Physics, Ohio University, Athens, OH 45701}
\author{K.A. Chipps}
\affiliation{Physics Division, Oak Ridge National Laboratory, Oak Ridge, TN 37831}
\affiliation{Department of Physics and Astronomy, University of Tennessee, Knoxville, TN 37996}
\author{M. Couder}
\affiliation{Department of Physics and Astronomy, University of Notre Dame, Notre Dame, IN 46556}
\author{ C. Deibel}
\affiliation{Department of Physics and Astronomy, Louisiana State University, Baton Rouge, LA 70803}
\author{A. Hood}
\affiliation{Department of Physics and Astronomy, Louisiana State University, Baton Rouge, LA 70803}
\author{M. Horana Gamage}
\affiliation{Department of Physics, Central Michigan University, Mt. Pleasant, MI 48859}
\author{R. Jain}
\affiliation{Department of Physics and Astronomy, Michigan State University, East Lansing, MI 48823}
\affiliation{Facility for Rare Isotope Beams, Michigan State University, East Lansing, MI 48823}
\author{C. Maher}
\affiliation{Department of Physics and Astronomy, Michigan State University, East Lansing, MI 48823}
\affiliation{Facility for Rare Isotope Beams, Michigan State University, East Lansing, MI 48823}
\author{S. Miskovitch}
\affiliation{Department of Physics and Astronomy, Michigan State University, East Lansing, MI 48823}
\affiliation{Facility for Rare Isotope Beams, Michigan State University, East Lansing, MI 48823}
\author{J. Pereira}
\affiliation{Facility for Rare Isotope Beams, Michigan State University, East Lansing, MI 48823}
\author{T. Ruland}
\affiliation{Department of Physics and Astronomy, Louisiana State University, Baton Rouge, LA 70803}
\author{M.S. Smith}
\affiliation{Physics Division, Oak Ridge National Laboratory, Oak Ridge, TN 37831}
\author{M. Smith}
\affiliation{Department of Physics and Astronomy, Michigan State University, East Lansing, MI 48823}
\affiliation{Facility for Rare Isotope Beams, Michigan State University, East Lansing, MI 48823}
\author{I. Sultana}
\affiliation{Department of Physics, Central Michigan University, Mt. Pleasant, MI 48859}
\author{C. Tinson}
\affiliation{Department of Physics and Astronomy, Michigan State University, East Lansing, MI 48823}
\affiliation{Facility for Rare Isotope Beams, Michigan State University, East Lansing, MI 48823}
\author{A. Tsantiri}
\affiliation{Department of Physics and Astronomy, Michigan State University, East Lansing, MI 48823}
\affiliation{Facility for Rare Isotope Beams, Michigan State University, East Lansing, MI 48823}
\author{A. Villari}
\affiliation{Facility for Rare Isotope Beams, Michigan State University, East Lansing, MI 48823}
\author{L. Wagner}
\affiliation{TRIUMF, 4004 Wesbrook Mall, Vancouver, BC, Canada V6T 2A3}
\author{R.G.T. Zegers}
\affiliation{Department of Physics and Astronomy, Michigan State University, East Lansing, MI 48823}
\affiliation{Facility for Rare Isotope Beams, Michigan State University, East Lansing, MI 48823}

\date{\today}

\begin{abstract}
The synthesis of heavy elements in supernovae is affected by low-energy (n,p) and (p,n) reactions on unstable nuclei, yet experimental data on such reaction rates are scarce. The SECAR (SEparator for CApture Reactions) recoil separator at FRIB (Facility for Rare Isotope Beams) was originally designed to measure astrophysical reactions that change the mass of a nucleus significantly. We used a novel approach that integrates machine learning with ion-optical simulations to find an ion-optical solution for the separator that enables the measurement of (p,n) reactions, despite the reaction leaving the mass of the nucleus nearly unchanged. A new measurement of the $^{58}$Fe(p,n)$^{58}$Co reaction in inverse kinematics with a 3.66$\pm$0.12 MeV/nucleon $^{58}$Fe beam (corresponding to 3.69$\pm$0.12 MeV proton energy in normal kinematics) yielded a cross-section of 20.3$\pm$6.3 mb and served as a proof of principle experiment for the new technique demonstrating its effectiveness in achieving the required performance criteria. This novel approach paves the way for studying astrophysically important (p,n) reactions on unstable nuclei produced at FRIB.
\end{abstract}

\keywords{machine-learning, recoil separators, ion-optics, nuclear reactions, nuclear astrophysics}

\maketitle

\section{\label{sec:intro}Introduction}
In core-collapse supernovae, the neutrino-driven wind off the nascent proto-neutron star has long been proposed as a potential site for heavy-element nucleosynthesis beyond iron \cite{Arcones2013}. Recent astrophysical model studies of the neutrino-driven wind environment indeed indicate that most winds are proton-rich for the majority of their duration \cite{wangNeutrinodrivenWindsThreedimensional2023a, pascalProtoneutronStarEvolution2022, fischerNeutrinoSignalProtoneutron2020}. In that case, the wind may harbor a $\nu$p-process that could explain the origin of the large abundances of neutron-deficient isotopes of Mo and Ru present in the solar system \cite{frohlichNeutrinoInducedNucleosynthesis642006a, pruetNucleosynthesisEarlySupernova2006, wanajoRpProcessNeutrinodrivenWinds2006}. The $\nu$p-process produces heavy elements via a sequence of rapid proton captures and (n,p) reactions~\cite{frohlichNeutrinoInducedNucleosynthesis642006a}. The process starts with seed nuclei in the iron region and continues on the proton-rich side of the valley of stability, close to the proton drip line. The (n,p) reactions involve unstable nuclei and are enabled by a small fraction of neutrons, which is maintained through neutrino interactions with the abundant protons. These interactions bridge the comparatively slow $\beta$-decays that would otherwise be required to produce heavy elements. Accurate rates of the (n,p) reactions are needed to predict the nucleosynthesis outcomes. Sensitivity studies have shown that even a deviation of a factor of 2 in some reaction rates can significantly affect the abundances of heavier elements \cite{Wanajo_Janka_2011, mnras2019}. Low energy (n,p) and (p,n) reactions on unstable nuclei have also been theorized to play an important role during explosive silicon burning in core-collapse supernovae \cite{woosleyExplosiveBurningOxygen1973, theNuclearReactionsGoverning1998, magkotsiosTRENDS44Ti56Ni2010a, subediSensitivity44Ti56Ni2020, hermansenReactionRateSensitivity2020}. These reactions are part of the complex nuclear reaction network in the silicon-iron region, which is especially significant for understanding the synthesis of long-lived radioactive isotopes. In particular, both types of reactions affect the production of radioactive $^{44}$Ti, which has been detected through its $\gamma$- and X-ray radiation in supernova remnants \cite{borkowskiRadioactiveScandiumYoungest2010, grebenevHardXrayEmissionLines2012, grefenstetteAsymmetriesCorecollapseSupernovae2014}, and via its decay products in meteorites~\cite{nittlerAstrophysicsExtraterrestrialMaterials2016}. They also influence the synthesis of a broad range of additional $\gamma$-ray emitters potentially observable in future galactic supernovae \cite{timmesCatchingElementFormation2019, hermansenReactionRateSensitivity2020}. Thus, reliable nuclear reaction rates for (p,n) are important to enable the interpretation of such observations via supernova explosion models \cite{andrewsNucleosyntheticYieldsCorecollapse2020}, especially to prepare for future $\gamma$-ray observatories such as the planned COSI mission~\cite{COSI}. 

The astrophysical (n,p) and (p,n) reaction rates are related to each other via the detailed-balance principle, and therefore, measurements constraining the rates can be performed in either direction \cite{Gastis2020a, Gastis2020b}. Nevertheless, experimental data on unstable nuclei for these reactions are extremely limited. Recently, the $\nu$p-process reaction rate of $^{56}$Ni(n,p)$^{56}$Co was measured using a neutron beam impinging on a radioactive $^{56}$Ni sample \cite{leeBetterUnderstandingNup2023}. However, this technique is confined to species with sufficiently long half-lives to produce a target and thus, not applicable to the vast majority of reactions of astrophysical interest. Additionally, the target activity introduces a radiation background that can reduce detector sensitivity to transitions to higher excited states, presenting further challenges. In recent years, there have been attempts to use the surrogate ratio method for determining (n,p) and (n,xp) cross sections of unstable nuclei, although these measurements have primarily targeted a higher energy range \cite{Gandhi_2022_surrogate_np} than what is relevant for the study of astrophysical processes.

In this paper, we present a novel technique for low-energy (p,n) studies on unstable nuclei for astrophysical purposes. Using machine learning algorithms, we modified the ion optics of SECAR (SEparator for CApture Reactions) \cite{bergDesignSECARRecoil2018}, a new recoil separator constructed at the National Superconducting Cyclotron Laboratory/Facility for Rare Isotope Beams (NSCL/FRIB) at Michigan State University, to enable its use for (p,n) reaction measurements. The technique uses inverse kinematics, where the heavy beam impinges on a light target, similar to methods used at much higher energies to probe weak interactions \cite{SasanoPRL}. Measurements of the small astrophysical reaction cross sections (typically on the order of some tenths of millibarns or lower) require the subsequent separation of beam and reaction products \cite{Gastis2020b}. The difficulty in performing (p,n) reaction measurements with a mass separator stems from the substantial similarity in mass between the unreacted beam ions and the heavy reaction product. To address this challenge, we utilized a multi-objective evolutionary algorithm~\cite{zhangMOEAMultiobjectiveEvolutionary2007} to explore the complicated parameter space, associated with individually adjusting the ion-optical elements of SECAR and ultimately identifying a solution that enables (p,n) reaction measurements. This shows that modern machine-learning approaches can facilitate the development of innovative applications of ion optical systems extending well beyond their original design. We demonstrate this new capability through the direct measurement of the $^{58}$Fe(p,n)$^{58}$Co reaction using a stable $^{58}$Fe beam. The technique shows promise for broad application to other beam species, including reaccelerated radioactive beams from FRIB.

The $^{58}$Fe(p,n)$^{58}$Co reaction is part of the explosive silicon-burning nuclear reaction network in core-collapse supernovae. Whilst serving as an important test case for statistical model predictions that are employed for the vast majority of silicon burning and $\nu$p-process reaction rates, it also has direct relevance for the synthesis of $^{59}$Fe, a long-lived $\gamma$-ray emitter that may be detectable in a future galactic supernova \cite{timmesCatchingElementFormation2019}. $^{58}$Fe(p,$\gamma$)$^{59}$Co has been identified as being important for the production of $^{59}$Fe \cite{hermansenReactionRateSensitivity2020}, but the competition with the (p,n) channel must be understood as well. Current statistical models predict a cross-over of (p,n) and (p,$\gamma$) cross sections at around 4.5~GK, which is in the relevant temperature range for supernova nucleosynthesis. 

There have been four previous measurements of the $^{58}$Fe(p,n)$^{58}$Co reaction cross section in the relevant low energy range below 4 MeV proton energy. The two measurements by Tims et al. (1993)~\cite{timsCrossSectionsReactions1993b} utilized a step-wise activation technique and direct neutron detection, respectively. The step-wise activation method is mostly insensitive to the population of the 25~keV $^{58}$Co isomer, as its decay is not rapid enough; therefore, it relies on a correction factor derived from statistical model calculations to determine the total cross section. For the assumed relatively small correction of 1-4\% the results from the two methods are in good agreement. A subsequent measurement by Sudar et al. (1994)~\cite{sudarExcitationFunctionsProton1994a} of the total cross section for the ground and isomeric states, performed via a single activation in a foil stack, yielded a slightly lower value in the energy region of interest. Additionally, a more recent measurement by Gosh et al. (2017)~\cite{GHOSH201786} using a similar activation technique, reported a significantly higher value, approximately four times greater than that of Tims et al. (1993) ~\cite{timsCrossSectionsReactions1993b}. The measurement of only the isomeric state by Sudar et al. (1996)~\cite{sudar_isomeric_1996} suggests a similarly large total cross section based on reasonably predicted population fractions of the isomer. Given the significant discrepancies among previous measurements, a re-measurement of the low-energy cross section using a completely different approach, as presented here, is important for advancing astrophysical reaction networks and testing statistical models.

\section{\label{sec:optics}Ion Optics Design} 
\begin{figure}[hb]
\includegraphics[width=\linewidth]{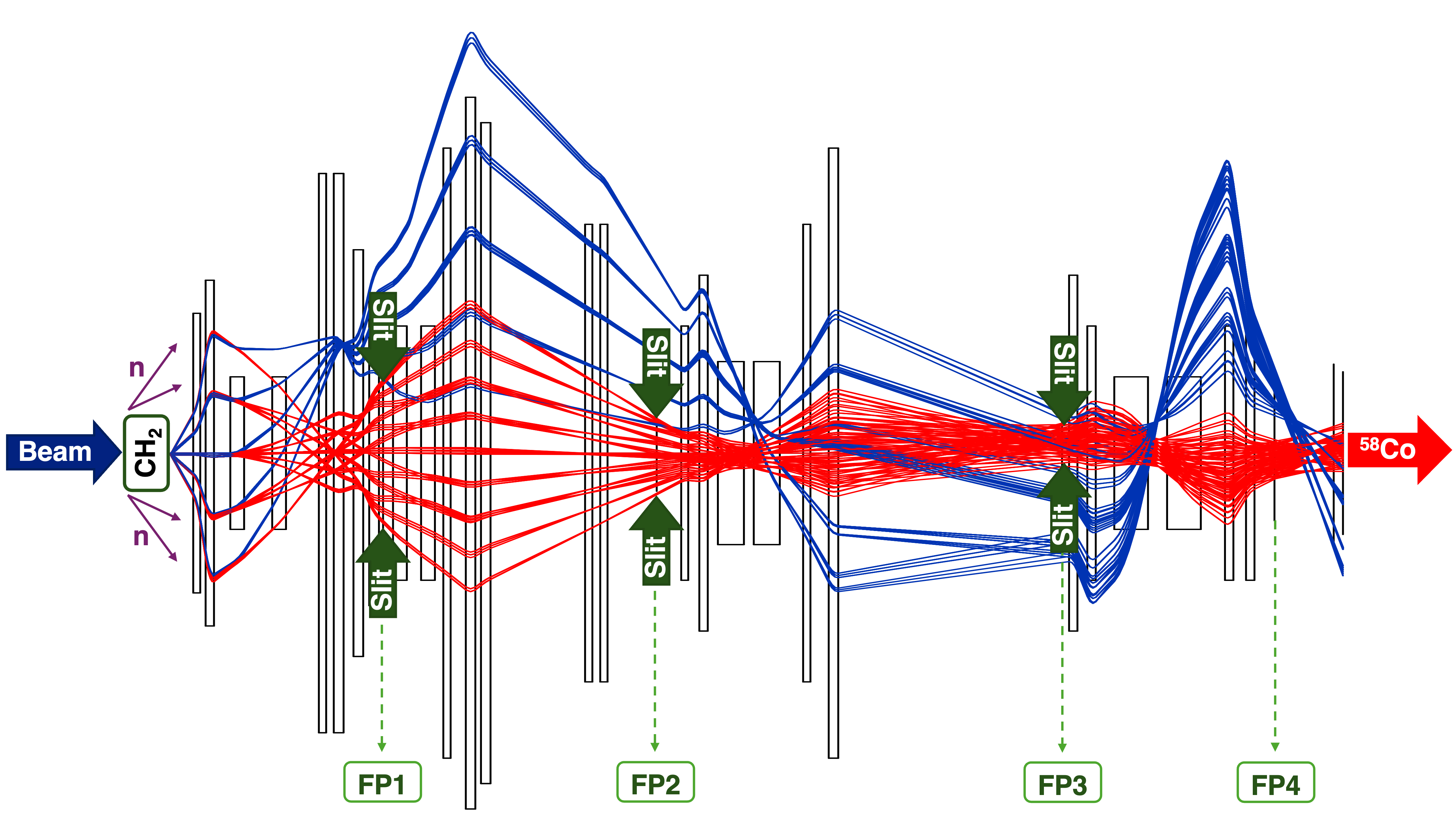}
\caption{\label{Fig:optics} New SECAR ion optics for (p,n) reaction measurements developed with a multi-objective evolutionary algorithm MOEA/D. The red and blue trajectories indicate the transport of reaction recoil and beam particle distributions with charge 22+ on the horizontal (xz) plane, respectively. Distributions correspond to an energy spread of $\pm$1.5\% and an angular spread of 20 mrad. Slits located at the original focal plane positions are indicated with vertical arrows and are inserted to block part of the beam ions.}
\end{figure}

The SECAR separator consists of 8 dipole magnets, 15 quadrupole magnets, 3 hexapole magnets, 1 octupole magnet, and 2 velocity filters and is optimized for radiative capture reaction measurements~\cite{bergDesignSECARRecoil2018} in inverse kinematics, where beam and heavy reaction products are separated by mass. The recoil separator sufficiently suppresses the beam particles to allow detectors at the final focus to resolve and uniquely identify recoil events, which may then be tallied to determine the reaction yield from which the cross section can be calculated. Suppression of unwanted beam particles is accomplished with slits at three locations (FP1, FP2, FP3) and the detectors downstream of FP4 (Figure~\ref{Fig:optics}). 

Since the standard magnetic and electrostatic field settings separate particles by mass, they are not applicable for (p,n) reaction measurements, where the beam and recoil have almost the same mass. We, therefore, used a Multiobjective Optimization Evolution Algorithm based on Decomposition (MOEA/D) \cite{zhangMOEAMultiobjectiveEvolutionary2007} method, specifically using the Tchebycheff decomposition approach, to search for a new combination of these fields that provides separation for (p,n) reactions' products, while retaining the system's existing geometric configuration. This optimization approach can accommodate a broad range of objectives that need to be optimized simultaneously. The method ensures diversity in solutions by employing differential evolution (DE) to generate new candidate magnet settings and polynomial mutation to introduce variability at each generation.

The MOEA/D framework defines a ``neighborhood" of sub-problems based on the similarity of their weighting vectors (combinations of the different parameters), which encourages information sharing among closely related sub-problems to enhance search efficiency. By transforming the multi-objective problem into manageable single-objective tasks, these sub-problems are then optimized simultaneously, and their solutions are combined to form the Pareto Front~\cite{ParetoMOEA}, which represents the optimal trade-offs between the potentially conflicting objectives. 

The entire multi-objective problem is conceptualized here as coupling the parameter space of 19 electromagnetic elements to the objective space of desired beam characteristics. These objectives included the maximization of the physical separation of reaction recoil and beam at the locations of FP2 and FP3, so existing slits can be used to stop unwanted beam particles. Another objective was the minimization of the beam width at each ion-optical element and the focal plane detectors to ensure maximum recoil transmission and detection, respectively. 

To evaluate parameter choices (magnetic field settings), the algorithm was coupled to beam transport simulations using COSY-INFINITY \cite{makinoCOSYINFINITYVersion2006}. For a given set of magnet settings, COSY calculates the transport of beam and recoil particles through the system, providing expected angle and energy distributions at the target location (up to 20 mrad and $\pm$1.5\% energy offset with respect to the central ray). It then returns particle envelope sizes and positions at every location within SECAR. Hence, each set of magnet settings generated by MOEA/D required a COSY simulation to evaluate its fitness values. 

The problem is initialized with a population of 1000 points, whose magnet settings were randomly selected from a sufficiently large parameter space. This initial population is then evolved over several hundred generations in order to fully explore the parameter space, converge on global rather than local minima, and identify Pareto optimal solutions. A Pareto Front~\cite{ParetoMOEA} represents the set of solutions where no single objective can be improved without worsening another, providing a trade-off between competing objectives. Every individual combination of parameters and objectives that forms part of this trade-off is included in the final set of solutions from which the Pareto Front is constructed. Several of these solutions were then selected for further evaluation with beam measurements. 

The configuration that was chosen had the ion-optical foci moved downstream from the original FP2 and FP3 locations to create separation between the beam and recoil at the slit positions (Figure~\ref{Fig:optics}) and did not use the velocity filters. The beam rejection of the separator is defined as the ratio of the detected beam particles, after accounting for the efficiency of the final focal plane detectors, over the total beam ions incident on the target. By carefully setting all three slit systems, a rejection rate of \EE{2}{-3} using only the separator was achieved, which was adequate for conducting the experiment.  

\section{\label{sec:setup}Experimental Approach}
The $^{58}$Fe(p,n)$^{58}$Co reaction measurement was performed at the NSCL facility at Michigan State University. A \EE{3.11}{6}~pps beam composed of 76$\pm$5~\% $^{58}$Fe$^{21+}$ and about 24\% of $^{58}$Ni$^{21+}$, at the energy of 3.785~MeV/nucleon in the laboratory frame, was delivered by the ReA3 accelerator and impinged on a 0.36$\pm$0.03~mg/cm$^2$ polyethylene (C$_2$H$_4$) target. The target thickness was determined by measuring its area and mass with a high-precision scale, considering the nominal polyethylene density of 0.93 g/cm$^3$. The thickness was further verified by measuring the energy loss of the beam as it passed through the target. The measurement was performed by determining the displacement of the beam downstream of the first two calibrated dipole magnets with and without the target. This approach confirmed the target thickness, resulting in a measurement of 0.39~mg/cm$^2$ with a deviation of about 7\% from the initial value, which falls within the uncertainty range of the areal density measurement. The energy loss in the target and the beam energy spread resulted in a reaction energy range of 3.66$\pm$0.12~MeV/nucleon in the laboratory frame, corresponding to a proton energy range of 3.69$\pm$0.12~MeV in a standard kinematics proton-beam experiment.

The total number of $^{58}$Fe beam ions impinging on the target was determined via hourly Faraday cup (FC) beam current measurements. The average of these measurements was used to calculate the total beam current integral for each hourly run. The stability of the beam current and the uncertainty associated with this integration were evaluated by analyzing the statistical deviation of all FC measurements from the overall average during the experiment. This analysis indicated relatively stable beam conditions, with a 10\% standard deviation that was adopted as the relative uncertainty in the total number of $^{58}$Fe beam ions used for the cross-section calculation.

Neutrons from the reaction were detected by four organic liquid scintillators (EJ-301), arranged in an azimuthal symmetric ring approximately 21 cm downstream of the target, covering angles from 15$^{o}$ to 35$^{o}$ in the laboratory frame. The neutron interactions with the experimental apparatus along with the neutron detection efficiency and solid angle were simulated using Geant4~\cite{Geant2003}, and validated by comparison with measurements using a $^{252}$Cf neutron source placed at the target location. 
 
The $^{58}$Co recoil ions traversed through the separator were detected at the final focal plane of SECAR, using an ionization chamber (IC), filled with isobutane (C$_4$H$_{10}$) gas at a pressure of 6.73 kPa, and a 2 mm thick double-sided silicon strip detector (DSSD) with 32 vertical (front) and 32 horizontal (back) strips, each with a pitch of 2 mm, providing a total active area of 64 mm x 64 mm. The ions' energy loss and remaining total energy were recorded by the IC and DSSD respectively, and in coincidence with the detection of neutrons at the target location. 

Accurate determination of the reaction yield required applying correction factors to account for various experimental conditions. Specifically, as beam particles traverse through the target, they undergo various charge-changing processes, leading to a distribution of multiple charge states. Recoil separators like SECAR can only transmit charge states that fall within specific magnetic and electric rigidity limits, making the precise determination of the outgoing recoils' charge state fraction essential for accurate cross-section calculations. Since only the 22$^{+}$ charge state of the $^{58}$Co recoils is selected and subsequently detected, it is crucial to accurately determine its fraction. The $^{58}$Co charge state distribution (CSD) was indirectly determined from a measurement of the $^{58}$Fe beam CSD after it passed through the target. This measurement was carried out using the methodology detailed in \cite{marshallMeasurementChargeState2023}. By analyzing the momentum dispersion on the viewer downstream of the B2 dipole (directly upstream of FP1, Figure \ref{Fig:optics}), we were able to quantify the charge state fractions of the $^{58}$Fe beam and compare these with semi-empirical models. The measured $^{58}$Fe CSD was consistent with the predictions of the Shima et al. (1982) model \cite{shimaEmpiricalFormulaAverage1982} within the experimental uncertainties. Given the close similarity in atomic structure between $^{58}$Fe and $^{58}$Co, the Shima model provided a robust prediction for the $^{58}$Co charge state distribution, with the 22$^{+}$ state being predominant. A direct comparison of the experimental and model-derived CSD of the $^{58}$Fe beam alongside the predicted charge state fractions for $^{58}$Co is presented in Table \ref{tab:csd}. The charge state correction factor for $^{58}$Co, used in the cross-section calculation, was thus determined to be 30.9$\pm$1.2\%, incorporating a 4\% relative uncertainty based on the agreement between the measured $^{58}$Fe CSD and the model predictions.

\begin{table}[htbp]
\centering
    \begin{tabular}{c|   c   |c   c}
        \hspace{0.7cm}\multirow{3}{*}{Q$^{+}$}\hspace{0.5cm} & \hspace{0.2cm}\multirow{1}{*}{Experiment}\hspace{0.2cm} & \multicolumn{2}{c}{Model prediction}  \\
        & $^{58}$Fe[$\%$] &  \hspace{0.2cm}$^{58}$Fe[$\%$] &  \hspace{0.2cm}$^{58}$Co[$\%$] \\
        \hline
        19 & 4.1$\pm$0.5 & 3.5 & 2.7 \\
        20 & 13.5$\pm$0.6 & 14.3 & 11.1 \\
        21 & 29.2$\pm$0.9 & 29.7 & 25.2 \\
        22 & 32.2$\pm$1.2 & 31.0 & 30.9 \\
        23 & 16.4$\pm$0.3 & 16.3 & 20.6 \\
        24 & 4.1$\pm$0.4 & 4.3 & 7.4 \\
        \hline
    \end{tabular}
    \caption{Charge state distribution (CSD) of the $^{58}$Fe beam as measured experimentally and compared with model predictions. The table lists the experimental fractions of $^{58}$Fe charge states alongside the predicted fractions for both $^{58}$Fe and $^{58}$Co based on the semi-empirical model from Shima et al. \cite{shimaEmpiricalFormulaAverage1982}. Due to the close similarity in atomic structure between $^{58}$Fe and $^{58}$Co, the Shima model was employed to estimate the $^{58}$Co CSD, which confirmed that the 22$^+$ state is the most prevalent and hence the one selected for detection. Consequently, the charge state correction factor for $^{58}$Co was determined to be 30.9$\pm$1.2\%, incorporating a 4\% relative uncertainty based on the comparison between measured $^{58}$Fe CSD and its model prediction.}
    \label{tab:csd}
\end{table}

Another correction accounts for transmission losses, as the SECAR slits that prevented the transmission of unwanted beam ions, also hindered some recoil particles from reaching the final focal plane detection system. Transmission measurements using the $^{58}$Fe beam and scaling the separator to different rigidities yielded data for a maximum angular spread of 4~mrad (angular straggling of the beam through the target) and multiple beam energies. Additionally, transmission measurements were performed with a $^{241}$Am $\alpha$-source of known intensity placed at the target position and collimated with a 1.5-mm diameter circular aperture, corresponding to the observed beam spot size. Two datasets were obtained using variable-sized apertures downstream of the target, which limited the emission angles to $<$10 and $<$20~mrad, respectively. The source emitted $\alpha$-particles were measured with a silicon detector inserted between the last aperture and the entrance of SECAR, and compared with the ones that reached the DSSD at the final focal plane, for obtaining the system's transmission. The energy dependence of the transmission was determined by again scaling SECAR to different magnetic rigidities corresponding to an energy acceptance range of $\pm$3\%. 

\begin{figure}[ht]
\includegraphics[width=\linewidth]{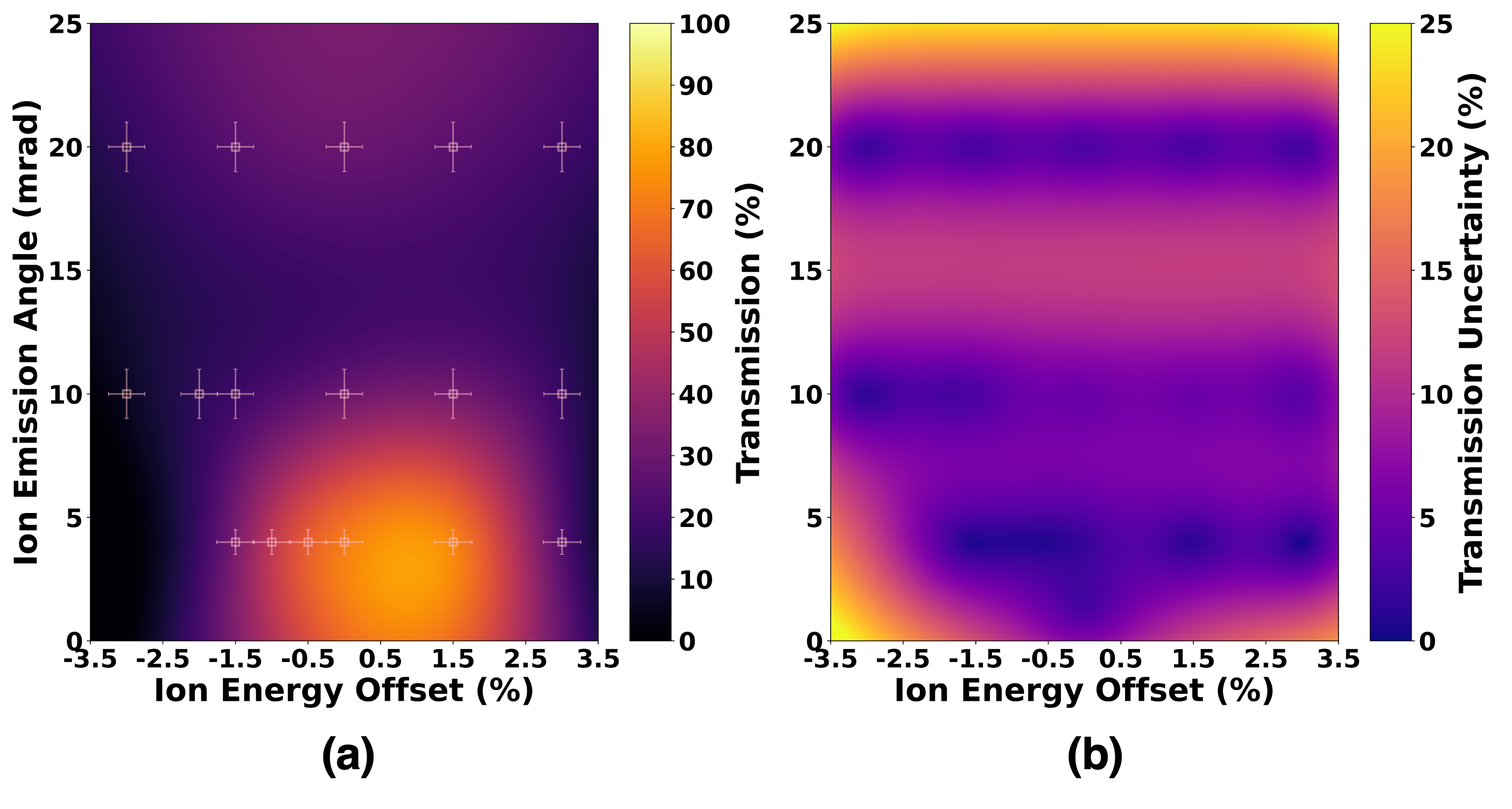}
\caption{\label{Fig:transmission} SECAR's transmission (a) and corresponding uncertainty (b) for the (p,n) optics as a function of emission angle and energy offset from the central ray. The system's transmission is derived, using a Gaussian process approach, from measurements with the $^{58}$Fe and $\alpha$-particles (white marks) at maximum angular spreads of 4~mrad, 10, and 20~mrad respectively.}
\end{figure}

The $^{58}$Co recoil distribution has a maximum angular spread of about 15~mrad and an energy spread of roughly $\pm$2\%. Since the transmission data obtained were unevenly distributed across these emission cone angles, interpolation was necessary to estimate values for angles without direct measurements. To achieve this, the collected transmission data were unfolded and analyzed, and then fitted using a Gaussian process (GP) method. This approach provided a continuous estimate of the system's transmission characteristics across the full range of emission angles and energies. The GP model was used with the Matérn kernel, selected for its ability to effectively handle both energy and angle inputs. The kernel's length-scale parameter, set to 1 for energy inputs and 2 for angle inputs, reflects the characteristic scales over which correlations in transmission vary with these parameters. Additionally, a smoothness parameter was set at 1.5 to balance the need to capture abrupt changes, such as those caused by the inserted slits, while also accounting for gradual transitions. This includes particles transmitted near the slits or those affected by the less intense parts of the beam profile. These parameters were specified as priors in the GP framework, reflecting initial assumptions about the transmission characteristics and their variability before incorporating observed data. This modeling resulted in an estimate of the system's transmission, along with its uncertainty, as a function of both energy and angle, shown in Figures~\ref{Fig:transmission}(a) and \ref{Fig:transmission}(b), respectively. While the data provides a general trend of the system’s performance, areas with sparse measurements exhibit higher uncertainty.

Both the total neutron detection efficiency and the recoil transmission depend on the reaction kinematics, which dictates the correlation between neutron and recoil angles and energies and depends on the excitation energy of the final state in $^{58}$Co. The relatively low incident beam energy allowed only the first 6 well-known excited levels of $^{58}$Co up to 457~keV to be populated. We estimated the relative population using the statistical model code TALYS~\cite{koningTALYSModelingNuclear2023}, indicating a dominant production of the ground state, accounting for around 50-60\% of the overall reaction yield, depending on the exact reaction energy within the target. In contrast, the third and fourth excited states were populated at about 25\% and 13\%, respectively. Less than 10\% of the reaction yield is expected to correspond to the remaining accessible states. The predicted neutron angular distributions were uniform in the center of mass. The resulting total transmission and total neutron efficiency averaged over all angles and energies, are 61.6$\pm$16.8~\% and 6.3$\pm$0.5~\%, respectively. Since the cross-section yield is determined by coincident detection of neutrons and heavy recoils, their combined detection efficiency must be considered. The system's overall efficiency is constrained by the energy-angle correlation of the reaction's products, transmission limits through SECAR, and the solid angle subtended by the neutron detectors. These factors lead to a restricted acceptance, which can be estimated through an event-by-event analysis. Accounting for neutron-recoil correlations, the average product of total neutron efficiency and heavy recoil transmission is 4.4$\pm$1.1 \%. The efficiency of the final focal plane detectors was measured to be a combined 95$\pm$2~\%, while the corresponding correction for data acquisition dead time was 10\%. 

\section{\label{sec:results}Results and Discussion}
\begin{figure}[htbp]
    \centering
    \includegraphics[width=\columnwidth]{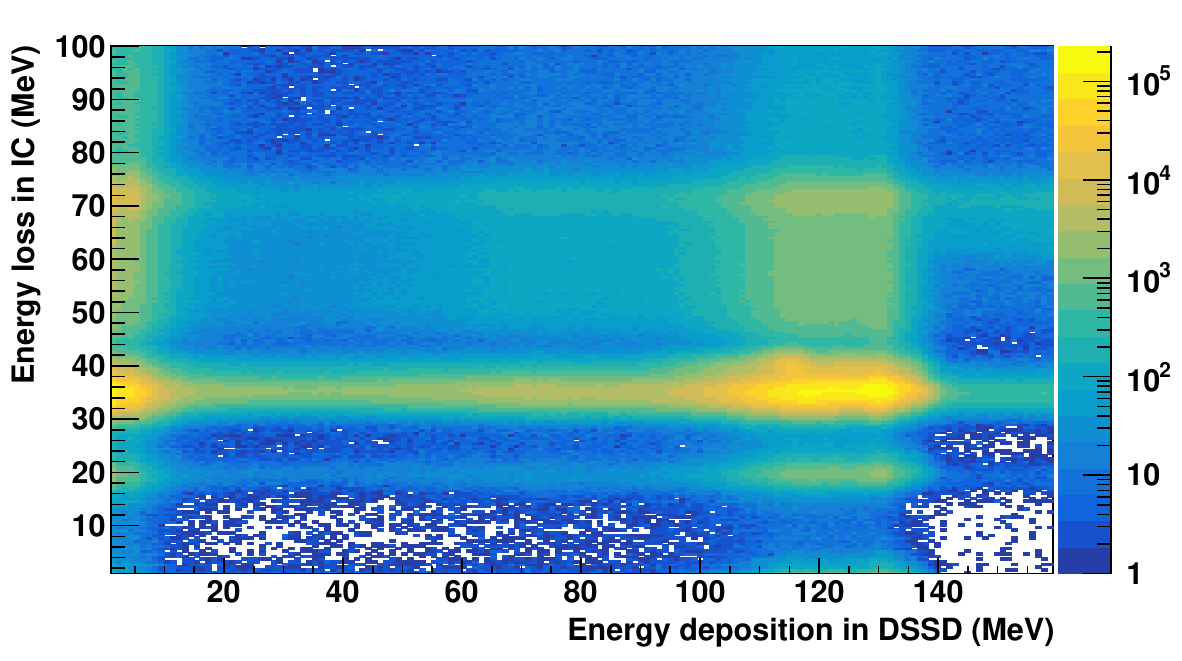}
    \caption{The energy loss of ions in the IC versus the remaining energy deposition in the DSSD defines the particle identification (PID) spectrum. The spectrum highlights the challenge of identifying recoil particles directly. To address this issue, in-coincidence detection methods are explored.}
    \label{fig:pid}
\end{figure}

\begin{figure}[htbp]
    \centering
    \includegraphics[width=\columnwidth]{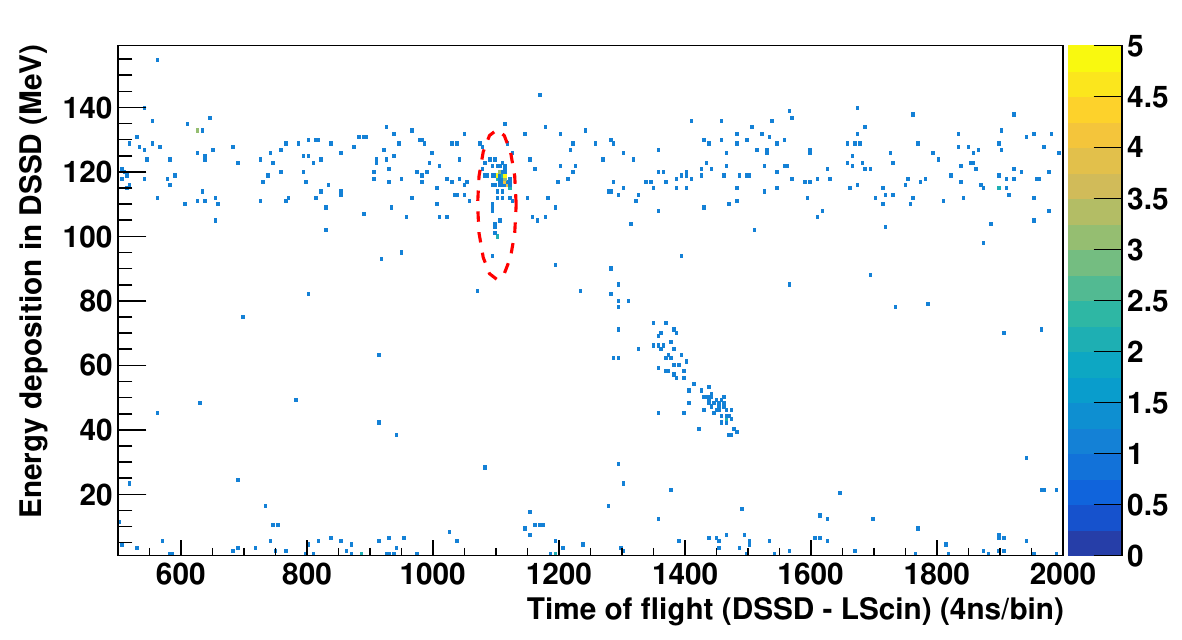}
    \caption{The energy deposition versus the time of flight difference between the neutron detection at the target location and the heavy ion detection at SECAR's final focal plane. The dashed red circle indicates the regions of interest for the (p,n) reaction recoils. Events occurring within the 1300–1500 ns time window and the energy range of 40 to approximately 80 MeV correspond to ions produced by fusion evaporation of the heavy ion beam on carbon.}
    \label{fig:tof_dssd_n}
\end{figure}

\begin{figure}[htbp]
    \centering
    \includegraphics[width=\columnwidth]{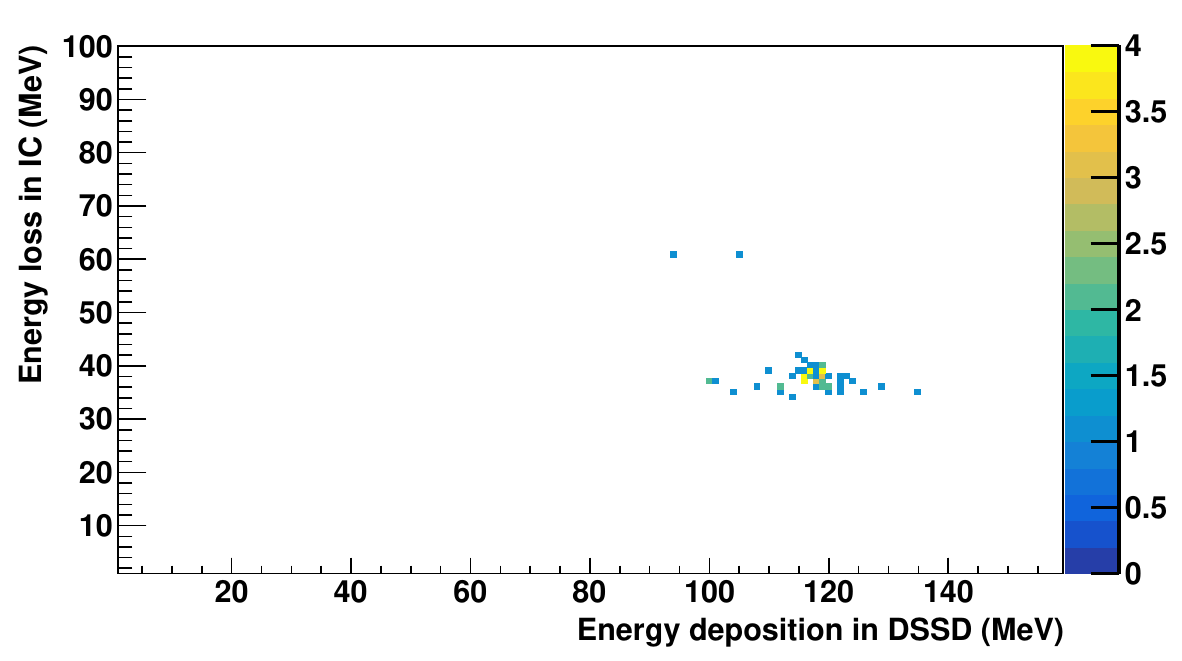}
    \caption{PID spectrum gated to detected neutrons at the target location within a 70 ns time window, shown within the red dashed circle in Figure \ref{fig:tof_dssd_n}. The integrated spectrum provides the recoil yield for cross-section calculations. Three random coincidences of neutrons with leaky beam events are observed at DSSD energies slightly higher than the recoil group ($>$125 MeV). Based on the leaky beam intensity distribution, no such events are expected in the recoil group, indicating that the combined rejection from the recoil separator and neutron tagging exceeds \EE{8}{-11}.}
    \label{fig:pid_n_tof}
\end{figure}

Figure~\ref{fig:pid} shows the particle identification (PID) spectrum at SECAR's detection plane. With neutron coincidence tagging and taking into account the 70 ns time difference between $^{58}$Co recoils at the final focal plane and neutrons near the target location, it was possible to clearly identify the (p,n) reaction recoils, as shown at the Figure~\ref{fig:pid_n_tof}. The underlying events in the time-of-flight (TOF) spectrum (Figure \ref{fig:tof_dssd_n}) arise from random coincidences between any ions reaching the DSSD and background events on the neutron detectors. Since these random events have no correlation, their timing signals are uniformly distributed within the 0.5 and 2 $\mu$s coincidence window, while the event group extending from 1300 to 1500 ns and at the energy range from 40 to roughly 80 MeV belongs to ions created from fusion evaporation on carbon. Figure~\ref{fig:pid_n_tof} shows the cleaned PID spectrum after applying an additional time-of-flight (TOF) gate within the 70 ns time window corresponding to the TOF of the recoils. In Figure~\ref{fig:pid_n_tof} there are three random coincidences of neutrons with leaky beam events, at DSSD energies slightly higher than the recoil group. Based on the leaky beam intensity distribution we do not expect any such events in the recoil group, implying a combined rejection from the recoil separator and neutron tagging surpassing \EE{8}{-11}. 

After accounting for charge-state fraction, recoil transmission, and detector efficiencies, a cross-section of 20.3$\pm$6.3~mb is obtained from the resulting reaction yield. The total relative uncertainty of 31\% arises from several key contributions. The largest source of uncertainty comes from systematic errors in recoil transmission and neutron efficiency, which together account for 62\%. Other significant contributions include beam intensity (10\%), inaccuracies in target thickness (6\%), and statistical uncertainty in the reaction yield (15\%). Additionally, beam purity measurement and charge state fraction calculation contribute modestly, at approximately 4\% and 2\%, respectively. These individual percentages collectively illustrate the various sources that comprise the total uncertainty (32\%).

\begin{figure}[ht]
\includegraphics[width=\linewidth]{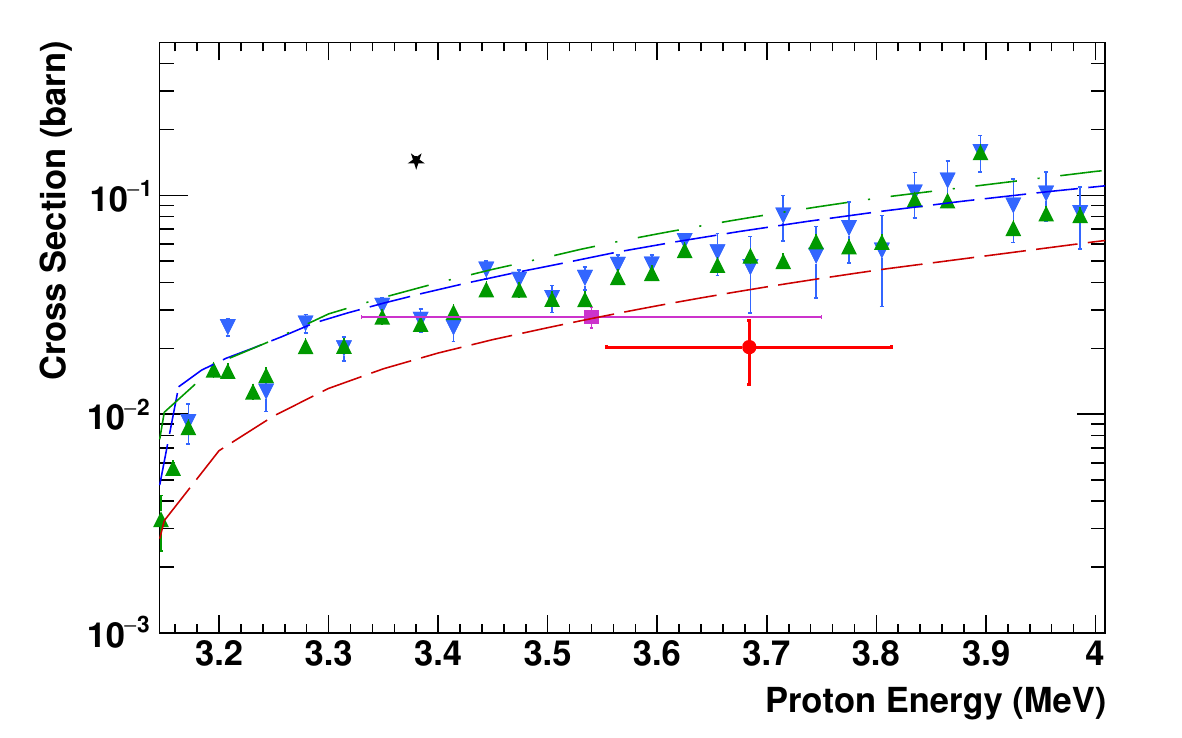}
\caption{\label{Fig:results} Cross sections for the $^{58}$Fe(p,n)$^{58}$Co reaction as functions of proton energy from previous measurements by \cite{timsCrossSectionsReactions1993b} (blue and green triangles), \cite{sudarExcitationFunctionsProton1994a} (purple square), \cite{GHOSH201786} (black star) and this work (red circle). All energies are given in proton energy in the laboratory frame considering normal kinematics. Also shown are statistical model predictions from TALYS using two different OMP, namely the Koning-Delaroche (green dash-dot-dash line), and the semi-microscopic JLM (dark red dashed lines) and from NON-SMOKER (blue dashed line).}
\end{figure}

The result of this work is shown in Figure~\ref{Fig:results} and in comparison with previous measurements and statistical model predictions obtained using TALYS~\cite{koningTALYSModelingNuclear2023} and NON-SMOKER~\cite{rauscherAstrophysicalReactionRates2000b}. Our measurement indicates a somewhat lower cross-section compared to previous experimental data, though the results agree with the measurements of Sudar et al. (1994)~\cite{sudarExcitationFunctionsProton1994a} and Tims et al. (1993)~\cite{timsCrossSectionsReactions1993b} within 2$\sigma$. We note that there are significant inconsistencies among the previous activation measurements. The measurement of the cross section for the $^{58}$Co isomeric state by Sudar et al. (1996) \cite{sudar_isomeric_1996} of 18 mb at 3.5~MeV would give a total cross section $>$120 mb for a reasonable isomer population fraction of $<$15\% and is under that assumption incompatible with Sudar et al. (1994) and Tims et al. (1993). Alternatively, taking the isomer production cross section of Sudar et al. (1996) and the Sudar et al. (1994) total cross sections at face value would result in an isomer production fraction of 65\% which is incompatible with statistical model predictions, and with the good agreement of the neutron and activation measurements of Tims et al. (1993). Indeed Sudar et al. (1996) do not provide an isomer fraction for 3.5~MeV indicating the authors themselves doubt their results. Additionally, Gosh et al. (2017)~\cite{GHOSH201786} reported a much higher cross-section overall at an even lower proton energy that is incompatible with both, the Sudar et al. (1994) and the Tims et al. (1993) activation measurements. The measurement by Tims et al. (1993) provides the most comprehensive data across the entire energy region of interest, obtained using both activation and direct neutron detection methods. However, it is important to note that for the activation result the cross section for the first isomeric state of $^{58}$Co$^{m}$ was not measured directly but inferred using statistical model predictions, that were significantly lower than predictions by Sudar et al. (1996), which appear to be in agreement with data at higher energies. In addition, the Tims et al. (1993) dataset depends on accurate estimation of beam energy losses through the target, particularly at lower energies. The associated systematic uncertainties were acknowledged but not included in the reported error bars.

Our results indicate that TALYS when using the optical model potential (OMP) of Koning-Delaroche~\cite{KONING2003}, shown in Figure \ref{Fig:results} with a blue dashed line, and NON-SMOKER~\cite{nonsmoker2001} (pink dashed line) both overestimate the cross section by about a factor of 3 compared to our measurement. In contrast, the TALYS prediction using the Jeukenne, Lejeune, and Mahaux (JLM) semi-microscopic OMP~\cite{JLM1977} results in lower cross-section values, with a difference of less than a factor of 2 from our results. It is also important to note that at such low energy and close to the reaction threshold, other parameters such as the gamma strength function or level density model have negligible impact on these predictions. Thus, the default TALYS parameters were used for these calculations, namely the “Simplified Modified Lorentzian (SMLO)” for the photon strength function and the “Gilbert-Cameron (GC)” model for the level density, respectively.

\section{\label{sec:conlcusions}Conclusions}

In summary, we have presented the first result obtained with the new SECAR recoil separator at NSCL/FRIB. We have taken advantage of a machine-learning approach to modify the ion optics and enable the use of SECAR for (p,n) reaction measurements. This broadens the separator's science reach significantly beyond the originally envisioned design scope. 

We applied this new technique to measure the $^{58}$Fe(n,p)$^{58}$Co reaction cross-section at a low energy close to the reaction threshold. Our results show a somewhat lower cross-section, but it remains within 2$\sigma$ of the previous measurements by ~\cite{timsCrossSectionsReactions1993b} and \cite{sudarExcitationFunctionsProton1994a}. Although performed at a slightly different energy, our findings do not support the significantly larger cross-sections reported more recently by \cite{GHOSH201786}. Similarly, the larger total cross sections, inferred from the relatively high isomeric state cross sections measured by \cite{sudar_isomeric_1996}, are also disfavored when considering reasonable population fractions for the $^{58}$Co isomeric state, ranging from 1-15\%, as predicted by statistical model calculations \cite{timsCrossSectionsReactions1993b,sudar_isomeric_1996}. 

This work builds upon initial work by \cite{Gastis2020b} using the ReA3 beam line magnets for a (p,n) measurement, that suggested a significant potential for improvements in efficiency and applicability when using a dedicated recoil separator. Specifically, the larger angular acceptance of SECAR (Figure \ref{Fig:transmission}), represents a substantial improvement over the about 8 mrad acceptance of the previous setup, while the energy acceptance is nearly doubled, increasing from about $\pm$1.5\% to $\pm$2.5\%. Additionally, SECAR achieved a maximum transmission efficiency of over 70\% for recoils, compared to just 20\% in the earlier experiment. These enhancements not only improve measurement efficiency but also overcome critical limitations of the previous configuration, which was confined to measuring a specific excited state of the nucleus rather than the complete (p,n) cross section. 

The method described in this paper represents a realization of the aforementioned advancements by combining machine-learning algorithms, ion-optical transport calculations, and experimental validation to optimize the SECAR ion-optical system for low-energy (p,n) measurements. It is an effective approach, anticipated to be applied to experiments with beams of unstable nuclei from FRIB's ReA3 accelerator, which can provide beam energies of 3-6~MeV/u, depending on the ion mass-to-charge ratio \cite{ReA_Leitner_2013, ReA_Villari_2022}. Considering these capabilities, the method opens up unique opportunities for ion-optical setups to perform direct low-energy (p,n) reaction measurements on unstable nuclei, offering critical insights into relevant astrophysical processes like explosive silicon burning and the $\nu$p-process.


\begin{acknowledgments}
This material is based upon work supported by the U.S. Department of Energy, Office of Science,
Nuclear Physics program under Award Numbers DE-SC-0022538 (CMU), DE-SC-0014384 (SECAR) and under contract DE-AC05-00OR22725 (ORNL), and grant number DE-FG02-88ER40387 (Ohio), and by the National Science Foundation (NSF) under award numbers PHY-1624942 (SECAR), PHY-1430152 (JINA-CEE), PHY-2209429, PHY-1102511 (NSCL) and has also benefited from NSF support under award number OISE-1927130 (IReNA).
This research used resources from the Facility for Rare Isotope Beams, which is a DOE Office of Science User Facility. We would like to thank Prof. Patrick O'Malley from the University of Notre Dame for providing the target foil used in this works
\end{acknowledgments}

\bibliographystyle{apsrev4-2}
\bibliography{secar_pn}

\end{document}